\documentclass[usegraphicx]{mn2e}
\usepackage[totalwidth=500pt,totalheight=680pt]{geometry}
\usepackage{times}
\newlength{\colwidth}
\title[AGN feedback in massive galaxies at $z\approx1$]%
{The prevalence of AGN feedback in massive galaxies at $z\approx1$}
\author[C.~Simpson et al.]{Chris Simpson$^1$\thanks{E-mail:
    C.J.Simpson@ljmu.ac.uk}, Paul Westoby$^1$,
Vinod Arumugam$^2$, Rob Ivison$^{2,3}$, \newauthor
Will Hartley$^{4,5}$ and Omar Almaini$^4$\\
$^1$Astrophysics Research Institute, Liverpool John Moores University,
Twelve Quays House, Egerton Wharf, Birkenhead CH41~1LD\\
$^2$Institute for Astronomy, University of Edinburgh, Royal Observatory,
Blackford Hill, Edinburgh EH9~3HJ\\
$^3$UK Astronomy Technology Centre, Science and Technology Facilities
Council, Royal Observatory, Blackford Hill, Edinburgh EH9~3HJ\\
$^4$School of Physics and Astronomy, University of Nottingham,
University Park, Nottingham NG7~2RD\\
$^5$Institute for Astronomy, ETH Z\"{u}rich, Wolfgang-Pauli-strasse
27, 8093 Z\"{u}rich, Switzerland}
\begin{document}

\date{Version of \today}

\pagerange{\pageref{firstpage}--\pageref{lastpage}} \pubyear{2012}

\maketitle

\label{firstpage}

\begin{abstract}
  We use the optical--infrared imaging in the UKIDSS Ultra Deep Survey
  field, in combination with the new deep radio map of Arumugam et
  al., to calculate the distribution of radio luminosities among
  galaxies as a function of stellar mass in two redshift bins across
  the interval $0.4<z\le1.2$. This is done with the use of a new
  Bayesian method to classify stars and galaxies in surveys with
  multi-band photometry, and to derive photometric redshifts and
  stellar masses for those galaxies. We compare the distribution to
  that observed locally and find agreement if we consider only objects
  believed to be weak-lined radio-loud galaxies. Since the local
  distribution is believed to be the result of an energy balance
  between radiative cooling of the gaseous halo and mechanical AGN
  heating, we infer that this balance was also present as long ago as
  $z\approx1$.  This supports the existence of a direct link between
  the presence of a low-luminosity ('hot-mode') radio-loud active
  galactic nucleus and the absence of ongoing star formation.
\end{abstract}

\begin{keywords}
galaxies: active --- galaxies: distances and redshifts --- galaxies:
evolution --- radio continuum: galaxies --- surveys
\end{keywords}

\section{Introduction}

The earliest radio surveys, such as the 3C catalogue of the northern
hemisphere (Bennett 1962; Laing, Riley \& Longair 1983), were
dominated by objects that we now classify as radio-loud active
galactic nuclei (AGN), extending out to unprecedentedly high
redshifts. Indeed, for most of the latter half of the twentieth
century, the object with the largest measured spectroscopic redshift
had originally been identified from a radio survey, due to the
very strong cosmic evolution of this population and the ease with
which such redshifts could be measured as a result of the strong
emission lines displayed by these powerful objects.

Modern radio maps now reach more than five orders of magnitude deeper
than the 3C catalogue with enormously better angular resolution and,
while these maps currently cover only a few square degrees, the Low
Frequency Array (LOFAR; Vermeulen 2012) and later the Square Kilometre
Array (SKA) will observe entire hemispheres to such
depths. Unsurprisingly, as these dramatic improvements in sensitivity
and resolutions have occurred, the nature of the new radio sources
that have been discovered has changed. At 1.4-GHz flux densities below
1\,mJy, the so-called microJansky population is known to comprise
significant numbers of star-forming galaxies (Condon 1984; Windhorst
et al.\ 1985, 1987) and radio-quiet AGN (Simpson et al.\ 2006a,
hereafter S06; Smol\v{c}i\'{c} et al.\ 2008; Ibar et al.\ 2009) that
are nearly absent from the original bright catalogues. This is not
solely due to the fact that deeper surveys probe lower luminosities at
a given redshift, but is driven by the very strong evolution in these
populations (Padovani et al.\ 2011; Simpson et al.\ 2012, hereafter
S12). By contrast, the most luminous radio sources, which display
similarly strong cosmic evolution, are too luminous to contribute at
sub-mJy flux densities even if they lie at extremely high redshifts.
Therefore the only radio-loud objects found at microJansky flux
densities are the low-luminosity radio galaxies whose weak evolution
out to at least $z\sim1$ (Dunlop \& Peacock 1990; Clewley \& Jarvis
2004; McAlpine \& Jarvis 2011) means they become a minority
population.

The evolution of these objects at $z\ga1$ remains uncertain, however.
Unlike the other populations, they are not readily identified at other
wavelengths (e.g., X-rays or the mid-infrared) and they do not possess
the strong emission lines that make redshift determinations easy. This
is believed to be because they are powered by `hot-mode' accretion
where material falls directly onto the supermassive black hole, and
hence they do not possess an accretion disc that is the source of
ionizing radiation in `cold-mode' sources. Some studies have reported
a decline in the space density of these objects (Waddington et al.\
2001; Rigby et al.\ 2011; S12) although they disagree
quantitatively. This is an issue for models that attempt to predict
the sky as seen by future radio telescopes since, while this
population is subdominant, it is still believed to contribute at the
$\sim20$\,per cent level (Wilman et al.\ 2008). Furthermore, these
objects are believed to play an important role in the evolution of
their host galaxies since the expanding radio source can evacuate
cavities in the intracluster medium and prevent gas from cooling and
fuelling further star formation, a phenomenon that has been directly
observed in X-ray images (e.g.\ B\"{o}hringer et al.\ 1993; Fabian et
al.\ 2003). The inclusion of this process in semi-analytic models of
galaxy formation, admittedly via a simplistic prescription, enabled
the steep cut-off at the bright end of the galaxy luminosity function
to be successfully reproduced (Croton et al.\ 2006; Bower et al.\
2006; Cattaneo et al.\ 2006).

In an effort to quantify the energetics of this phenomenon, termed
`radio-mode feedback' by Croton et al.\ (2006), Best et al.\ (2005a,b)
determined the distribution of radio luminosities for such objects at
$z<0.1$ using a combination of data from the Sloan Digital Sky Survey
(SDSS; York et al.\ 2000), the National Radio Astronomy Observatory
(NRAO) Very Large Array (VLA) Sky Survey (NVSS; Condon et al.\ 1998)
and the Faint Images of the Radio Sky at Twenty Centimetres (FIRST;
Becker, White \& Helfand 1995) survey. They found that hot-mode
radio-loud AGN were more prevalent in more massive galaxies and
suggested that this represented a duty cycle whereby the most massive
galaxies experienced AGN activity more frequently, \textbf{a
  conclusion also drawn by Shabala et al.\ (2008) from an analysis of
  the same data}. In a later paper, Best et al.\ (2006) estimated the
time-averaged heating luminosity for galaxies of different stellar
masses, and demonstrated that this heating was able to counterbalance
the gas cooling in early-type galaxies at all stellar masses,
supporting the idea that AGN feedback of this type is responsible for
limiting the stellar mass of galaxies.

Since the shape of the galaxy luminosity function is easily reproduced
with any model that suppresses gas cooling in massive haloes,
semi-analytic models are limited in their ability to constrain the
details of hot-mode feedback. Nevertheless, Bower, McCarthy, \& Benson
(2008) implemented additions to the Bower et al.\ (2006) model to
investigate the effect of feedback on the intracluster medium and
claimed that the energy deposited could expel gas from the central
regions of the halo and thereby explain the observed
luminosity--temperature ($L_X$--$T_X$) relation and baryon fractions
in groups and clusters of galaxies.  However, these additions
required, in the absence of other changes, delaying the onset of
hot-mode feedback (by increasing the mass at which a halo entered
hydrostatic equilibrium) to maintain agreement with the observed
galaxy luminosity function.

Higher-resolution hydrodynamical simulations can be expected to better
constrain this phenomenon, and these have incorporated supermassive
black holes in a variety of ways, with different prescriptions for AGN
feedback and different triggers for the transition between cold-mode
and hot-mode accretion. Sijacki et al.\ (2007), for example, employ
alternative models for energy deposition according to the mode of
accretion, which they assume depends only on the accretion rate as a
fraction of the Eddington limit. At high rates (cold-mode accretion),
a fraction of the accreted rest-mass energy heats gas particles close
to the black hole while the energy liberated during hot-mode accretion
is deposited in bubble-like regions away from the nucleus to mimic the
results from X-ray observations. Furthermore, the fraction of energy
released is different in the two modes due to the different
efficiencies with which radiative and mechanical energy is coupled to
the surrounding gas. Sijacki et al.\ (2007) find that this model
reproduces a number of observed properties of galaxies and the
intracluster medium, and Puchwein, Sijacki, \& Springel (2008) use it
to demonstrate the importance of AGN feedback in explaining the
$L_X$--$T_X$ relation. McCarthy et al.\ (2010), on the other hand, are
also able to reproduce the observed $L_X$--$T_X$ relation and gas mass
fractions for galaxy groups and clusters using the Booth \& Schaye
(2009) model that contains only one feedback prescription, similar to
that of Sijacki et al.'s cold-mode but with a threefold higher
efficiency.

The fact that different simulations can successfully reproduce the
same observed properties is perhaps indicative of the simplicity of
the feedback mechanism -- rapid cooling of gas increases the black
hole accretion rate and hence the heating power, while overheating
starves the black hole of fuel -- but means the details remain
unclear. Purely analytic models have been used to investigate the
phenomenon but similarly suffer from the lack of a direct coupling
between the cooling of the gas on large scales and the accretion onto
the black hole at small scales. This can only be a small fraction of
the classical cooling flow rate and Pope (2011) has suggested that the
fraction decreases with increasing galaxy mass, from $\sim10^{-2}$ in
low-mass ellipticals to $\sim10^{-4}$ in brightest cluster
galaxies. Additional observational data are therefore required to
unlock these details and physical parameters driving this mode of
feedback, as we still lack basic results such as the epochs at which
the heating/cooling balance is achieved in galaxies of different
masses. In addition, a deeper understanding of this relationship could
allow simulations of galaxy formation to predict the evolution of this
population and its contribution to the faint radio source counts in
the absence of reliable observational data. This is very different
from the situation for cold-mode feedback, where the ease of
identifying such objects across all redshifts allows them to be used
as observational constraints for incorporating AGN activity in
semi-analytic models (e.g., Fanidakis et al.\ 2011).

In this paper, we attempt to answer this question by studying radio
sources in the Ultra Deep Survey (UDS) field of the UKIRT (United
Kingdom Infrared Telescope) Infrared Deep Sky Survey (UKIDSS; Lawrence
et al.\ 2007), which benefits from extremely deep near-infrared and
optical imaging, the latter through its original incarnation as the
Subaru/\textit{XMM-Newton\/} Deep Field (SXDF; Furusawa et
al.\ 2008). The remainder of this paper is formatted as follows. In
Section~2 we describe the optical, near-infrared, and radio data in
the UKIDSS UDS field that we use to perform our analysis, including a
deeper radio map than the one used by S06 and S12 (Arumugam et al., in
preparation). Section~3 presents a new Bayesian method for classifying
individual objects as either stars or galaxies, the derivation of
photometric redshifts and stellar masses, and the results of applying
this method to sources in the UDS. Section~4 then details our analysis
of the AGN fraction among galaxies in this field, following the work
of Best et al.\ (2005b), and a summary of the paper is presented in
Section~5. Throughout this paper we adopt a flat $\Lambda$CDM
cosmology with $H_0=70$\,km\,s$^{-1}$\,Mpc$^{-1}$ and $\Omega_\Lambda
\equiv 1-\Omega_{\rm m} = 0.73$ (Komatsu et al.\ 2011). All magnitudes
are on the AB system.

\section{Imaging data}

\subsection{Near-infrared imaging and catalogue}

The starting point for the galaxy catalogue used here was the Eighth
Data Release (DR8) of UKIDSS. UKIDSS consists of five separate surveys
and the Ultra Deep Survey (UDS) covers 0.77\,deg$^2$, with the DR8
imaging reaching depths of $J=24.9$, $H=24.2$, $K=24.6$ (5$\sigma$
depths in 2-arcsecond diameter apertures, estimated from the
r.m.s.\ deviations between a large number of randomly-placed
apertures). We use the same $K$-band-selected catalogue as Hartley et
al.\ (2013), produced by merging the catalogues from two separate runs
of the SExtractor software (Bertin \& Arnouts 1996), one of which was
designed to deblend sources and the other to identify extended
low-surface brightness galaxies. Photometry for all sources was
measured in 3-arcsecond diameter apertures in all three filters, as
was the total $K$-band magnitude (MAG\_BEST). Following Hartley et
al.\ (2013), only objects with a total magnitude $K<24.3$ were used in
the subsequent analysis.

\subsection{Other imaging data}

We use the same images in the \textit{uBVRiz\/} filters as S12, with
astrometric corrections between the UKIRT and Subaru imaging being
calculated in an identical manner. A correction was applied to the
$u$-band photometry as described in S12, to account for a difference
in the point spread function (psf).

We use images in channels 1 and 2 of \textit{Spitzer\/}'s Infrared
Array Camera (IRAC) taken as part of the Legacy Program SpUDS (PI:
J.~Dunlop) and use the same method as S12 to account for the broader
psf of these images. The SpUDS images are deeper than those used by
S12, but do not cover a small portion of the combined UDS/SXDF region;
in these cases we use the SWIRE photometry of S12.

Imaging from the Galaxy Evolution Explorer (GALEX) satellite exists in
the form of a deep exposure from program GI6\_005 (PI: S.~Salim) plus
shallower exposures from the Deep Imaging Survey. We combined these
images to provide relatively uniform coverage of the entire UDS field
and obtained photometric measurements for all sources in the
near-ultraviolet (NUV) band. However, it was found that use of these
data did not measurably improve the quality of our photometric
redshifts and caused problems for sources in more crowded regions due
to the much larger psf of the GALEX data. We therefore did not use
these data in our analysis.

\subsection{A deeper 1.4-GHz radio map}

A new 1.4-GHz radio map was produced by combining the Very Large Array
B- and C-array data of S06 with A-array observations made at the same
14 pointing centres, which include the data of Ivison et al.\ (2007).
The synthesized beam size of this map is $1.8\times 1.6$\,arcsec$^2$
and the noise in the map is less than 8\,$\mu$Jy\,beam$^{-1}$ in the
deepest region. The method by which this new map, and an associated
model of the effect of bandwidth smearing, were produced are described
in detail by Arumugam et al.\ (in preparation).

\subsubsection{Source extraction}

The SExtractor software package was used to identify sources in the
radio map. Sources which were clearly extended or comprised of
multiple components (including those identified as such by S06) were
removed from the source list and has their total flux densities
estimated by summing the signal in specifically defined apertures on
an object-by-object basis. Thirty-four sources were found to be
significantly extended and handled in this way.

The flux densities of the remaining sources were derived by fitting an
elliptical Gaussian to a small region of the map around each source,
providing full-width at half-maximum (FWHM) measurements along the
major and minor axes of $\theta_{\rm M}$ and $\theta_{\rm m}$,
respectively. A source was considered to be unresolved if
\begin{equation}
\frac{\theta_{\rm M} \theta_{\rm m}}{\theta^*_{\rm M} \theta^*_{\rm m}} < 
\frac{1}{b(\alpha,\delta)}
\end{equation}
where $\theta^*_{M}$ and $\theta^*_{m}$ are the major and minor axes
of the beam, and $b(\alpha,\delta)$ is the bandwidth smearing at the
source location. Unresolved sources were assigned a flux density of $S
= S_{\rm peak}/b(\alpha,\delta)$ where $S_{\rm peak}$ was derived from
a simple surface fit to the source.

A few sources were not accurately fitted by a single elliptical
Gaussian, including sources with a bright core and lower surface
brightness extended emission. Flux densities for these sources were
obtained by smoothing the radio map with a Gaussian five times
broader than the synthesized beam and using elliptical Gaussian fits.
The resulting catalogue comprises 1465 radio sources detected with a
signal-to-noise ratio ${\rm S/N}>5$ within the SXDF/UDS overlap
region.

\subsubsection{Completeness}

\begin{figure}
\resizebox{\hsize}{!}{\includegraphics{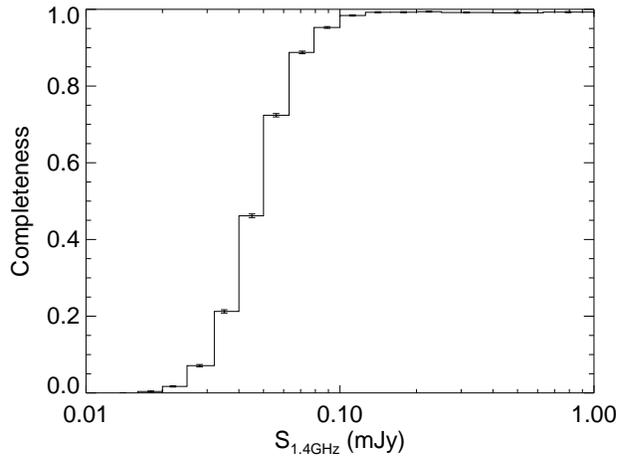}}
\caption[]{Completeness for radio sources in the UDS/SXDF overlap
  region, produced as described in the text.}
\label{fig:completeness}
\end{figure}

We undertake a new approach to estimating the completeness of the
radio map as a function of source flux, that takes into account both
the detectability of sources and the effects of Eddington bias
(Eddington 1913) simultaneously. We first produce a histogram of
source flux densities in bins of $\Delta\log S=0.1$, which represents
the distribution of real radio sources. In each of these bins we then
create 100 fake sources whose positions are uniformly distributed
across the UDS/SXDF overlap region, and whose flux densities follow
the source counts of Bondi et al.\ (2008; our results are not
significantly affected if we instead use the sixth-order polynomial
fit to the source counts from Hopkins et al.\ 2003). These are
injected into the radio map and sources extracted in the manner
described above. A new flux density histogram is produced and the
original histogram of real radio sources subtracted from it. Each
`difference histogram' is then scaled by the expected number of
sources in its input flux density range and all of them are added
together; this histogram represents the measured flux density
distribution for sources that follow the counts of Bondi et
al.\ (2008). Finally, the number of sources in each bin is divided by
the number of input sources in that bin to provide a histogram of the
effective completeness. This entire process is repeated 100 times to
provide a reliable estimate of the completeness as a function of flux
density, and the uncertainty in this estimate. The results are
presented in Fig.~\ref{fig:completeness}.

\subsubsection{Assigning near-infrared counterparts}
\label{sec:identifications}

Complex extended radio sources were assigned $K$-band counterparts by
visual inspection, generally following the original optical
identifications made by S06. Counterparts were assigned to other
sources using the criterion of the likelihood ratio (de Ruiter, Arp \&
Willis 1977; Wolstencroft et al.\ 1986) was used, as developed by
Sutherland \& Saunders (1992) and employed by S06. For each radio
source $r$, the likelihood ratio $L_{kr}$ is calculated for all
catalogued $K$-band sources within 5 arcseconds of the radio position
as follows:
\begin{equation}
L_{kr} = \frac{Q(<m_k) \exp (-r_{kr}^2/2)}{2\pi \sigma_x \sigma_y
  N(<m_k)} \, .
\end{equation}
In this equation, $Q(<m_k)$ is the fraction of radio sources whose
counterparts are brighter than the proposed identification, $\sigma_x$
and $\sigma_y$ are the uncertainties in the radio position, $r_{kr} =
\sqrt{(\Delta x/\sigma_x)^2 + (\Delta y/\sigma_y)^2}$ is the
normalized separation between the radio and infrared positions, and
$N(<m_k)$ is the true surface density of sources brighter than the
proposed identification. As in S06, $\sigma_x$ and $\sigma_y$ are
calculated by adding a quadrature a 0.2-arcsecond astrometric
uncertainty to the formula of Reid et al.\ (1988).

The probability that a particular $K$-band object is the correct
identification for a radio source is then given by
\begin{equation}
P_{kr} = \frac{L_{kr}}{\sum\limits_k L_{kr} + (1-Q)} \, ,
\end{equation}
where $Q$ is the fraction of radio sources with $K$-band counterparts,
which is estimated from the data to be 97\,per cent.

\section{Photometric redshifts and stellar masses}

We employ a novel three-step process to derive photometric redshifts
and stellar masses, and remove stars from our object catalogue. Before
undertaking this process, it is necessary to eliminate any photometric
measurements of objects that are saturated in the Suprime-Cam images.
Since we are interested in $z>1$ galaxies, which lie far below the
saturation limit, our analysis will not be affected if we err on the
side of caution and remove objects that are close to the saturation
limit. For each of the five Suprime-Cam filters we determine the
saturation limit by plotting the fluxes of all objects, measured in a
0.7-arcsecond diameter aperture, against the $u$-band flux in the same
aperture and identifying the pronounced flattening. The limits thus
derived are 11.5\,$\mu$Jy in \textit{BVRi\/} and 30\,$\mu$Jy in $z$,
and each saturated measurement is removed from the photometric
catalogue. Objects where four or five of the Suprime-Cam measurements
are saturated are eliminated from the catalogue.

\subsection{Photometric redshifts}

We have used the publicly-available code EAzY (Brammer, van Dokkum \&
Coppi 2008) to derive photometric redshifts for galaxies within the
UDS/SXDF overlap region. This code finds the best fit of a linear
combination of template spectra and initially we used the standard set
of six galaxy templates to derive photometric redshifts for the sample
of 3116 galaxies with reliable spectroscopic redshifts (i.e.,
spectroscopically-identified stars and quasars were excluded). This
resulted in a systematic overestimate of the redshifts for $z\sim2$
galaxies that could only be eliminated by applying an offset to the
$u$-band photometry of $-$0.3\,mag, far in excess of the true
zeropoint uncertainty. By comparing the spectra of these objects with
their template fits, we found that the spectra had redder rest-frame
ultraviolet--optical colours than the fits. At $z\sim2$, this results
in faint observed $u$-band fluxes that the templates cannot fit at the
true redshift without applying a large zeropoint offset, instead
preferring a systematically-higher redshift that pushes the Lyman
break further into the filter bandpass. However, while the application
of this offset improves the quality of the photometric redshifts at
$z\sim2$, it must inevitably reduce their quality in other redshift
ranges where the templates provide a good fit to the true photometry.

The need to apply photometric offsets larger than the zeropoint
uncertainties is indicative of an inadequate set of galaxy
templates. A corollary of this is that a suitable set of templates
\textit{must\/} provide reliable photometric redshifts without the
need for large zeropoint offsets. By understanding the reason for the
original failure, we were quickly able to produce a suitable template
set from the Bruzual \& Charlot (2003) simple stellar population
models (BC03). This set comprises six simple stellar population
templates with ages logarithmically spaced between 30\,Myr and
10\,Gyr, plus mildly-reddened ($A_V=0.25$\,mag) versions of the two
youngest templates and a more heavily-reddened ($A_V=1.0$\,mag)
version of the 30-Myr-old template. All spectra have Solar
metallicity. A Chabrier (2003) initial mass function (IMF) was adopted
and the Pei (1992) parametrization of the Small Magellanic Cloud (SMC)
extinction law was used. This template set did not produce any
large-scale deviations in the $z_{\rm phot}$--$z_{\rm spec}$ plane and
we further tested the suitability of the templates by varying the
photometric zeropoints in order to minimize the normalized median
absolute deviation $\sigma_{\rm NMAD}$ (Brammer et al.\ 2008). We
obtained $\sigma_{\rm NMAD}=0.027$ ($\sigma_{\rm NMAD}=0.023$ in the
redshift interval $0.4\leq z<1.2$ that we study in Section~4) with
offsets that were smaller in magnitude than 0.05\,mag in all filters,
and less than 0.03\,mag in most.

The absence of this effect in the test sample of Brammer et
al.\ (2008) is likely to be due to the optical magnitude limit
inherent in that sample, biasing it strongly in favour of galaxies
with blue rest-frame UV--optical colours. Over 70\,per cent of the
redshifts used by Brammer et al.\ come from the Chandra Deep
Field-South sample of Popesso et al.\ (2009), which has a magnitude
limit of $B<24.5$, whereas the UDSz and AGN samples that provide most
of our $z>2$ galaxies were selected with either no optical magnitude
limit, or a very faint limit. The blue templates themselves are likely
to be the result of the `inadequate [and] simplified treatment of dust
obscuration' in the synthetic galaxy sample from which Brammer et
al.\ (2008) derive their basis template set, possibly coupled with the
use of the Milky Way extinction law rather than the SMC law.

\subsection{Estimating stellar masses}

Stellar masses are derived for all sources by fitting a much larger
set of simple stellar population templates, keeping the redshift fixed
at the spectroscopic redshift, where available, or the best-fitting
photometric redshift. This set covers the same range of ages but at
much more closely-spaced intervals, and includes a larger set of young
reddened templates. At this stage we only fit the subset of templates
whose ages are less than the age of the Universe at the adopted
redshift, to ensure that the inferred star-formation history is valid.
This method allows us to recover any star-formation history
(cf.\ Reichardt et al.\ 2001) and hence current stellar mass (the
models are normalized to 1M$_\odot$ of formed stars, but we
renormalize them to 1M$_\odot$ of current stars, including
remnants). As a final step, the stellar mass derived from the
corrected aperture photometry is scaled by the ratio between the total
$K$-band flux and the aperture flux.

Although only Solar metallicity models are used to derive stellar
masses, the effect of using other metallicities is small. There are
systematic offsets, such that masses derived using subsolar models
($Z=0.4Z_\odot$) are larger while those derived using supersolar
models ($Z=2.5Z_\odot$) are smaller, but these are less than a factor
of two and, crucially, disappear for the massive galaxies
($M>10^{10.5}$\,M$_\odot$) being studied in Section~4.

\subsection{Star--galaxy separation}

Methods to identify stars in deep extragalactic surveys are many and
varied, often being based on morphological and/or colour criteria.
Morphological criteria are not trustworthy at faint magnitudes and
will misclassify objects of small intrinsic size, while colour
classification needlessly throws away much of the photometric
information and becomes unreliable when the signal-to-noise ratio in
one or more of the chosen filters is low. Our solution is to fit
stellar spectral energy distributions (SEDs) to the full photometric
data of each object and classify sources based on the relative quality
of the galaxy and stellar fits.

Since the stellar templates are fitted at a redshift of zero whereas
the redshift is a free parameter in the galaxy fits, the preferred
model is not simply the one that provides the lowest value of
$\chi^2$. Instead we employ a Bayesian analysis (e.g., Sivia 1996),
which produces the following result for the ratio of probabilities
between an object being a galaxy or star, $P_{\rm g}$ and $P_{\rm s}$,
respectively:
\begin{equation}
\frac{P_{\rm g}}{P_{\rm s}} = \frac{1-S}{S}
\frac{N_{\rm g}(K)}{N_{\rm s}(K)}
\frac{\int_{z_{\rm min}}^{z_{\rm max}} P(z|K) \exp (-\chi^2_{\rm g}/2)
  {\rm\,d}z}%
{\exp (-\chi^2_{\rm s}/2)} \, ,
\label{eqn:bayes}
\end{equation}
where $S$ is SExtractor's stellarity parameter, $z_{\rm min}=0.01$ and
$z_{\rm max}=7$ are the limits of the photometric redshift search, and
$P(z|K)$ is EAzY's prior redshift distribution for a galaxy of
magnitude $K$. The initial term indicates our prior expectation from
the relative sky density of galaxies (Gardner, Cowie \& Wainscoat
1993) and stars (from the model of Jarrett 1992) at each object's
$K$-band magnitude. We use the posterior probability density function
that EAzY outputs to numerically calculate the integral in the above
expression, and limit $S$ to lie in the range $0.005 \leq S \leq
0.995$.

The stellar SEDs we use are the photospheric models of Allard, Homeier
\& Freytag (2011), which cover a wide range of effective temperatures
and surface gravities. We also produce synthetic spectra for stellar
binaries by combining pairs of models that have been scaled by the
surface areas of each star, since the Allard et al.\ models give the
emission per unit area. We only consider main-sequence binaries and
use the mass--radius and mass--luminosity relations of Demircan \&
Kahraman (1991) to derive the effective temperatures, surface
gravities, and radii of main-sequence stars over the mass range $0.02
< M/{\rm M}_\odot < 30$ and interpolate within the grid of models. We
assume a constant stellar radius below 0.08M$_\odot$ (Chabrier et
al.\ 2008) and confirm that the mass--effective temperature relation
this implies is consistent with the measurements of Leggett et
al.\ (2010).

\begin{figure}
\resizebox{\hsize}{!}{\includegraphics{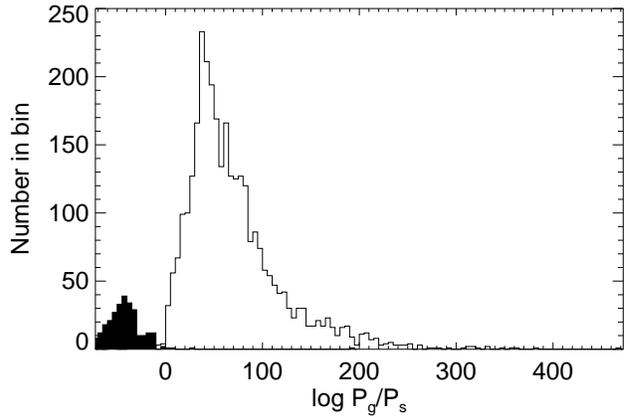}}
\caption[]{Histogram showing the ratio of probabilities between each
  object with a secure spectroscopic redshift being a galaxy or star,
  $P_{\rm g}/P_{\rm s}$. The filled histogram shows the ratio for
  spectroscopically-confirmed stars, while the open histogram shows
  the ratio for galaxies.}
\label{fig:bayes}
\end{figure}

We test our classification using the sample of 3421 objects with
reliable spectroscopic redshifts and at least two unsaturated
photometric measurements in the Suprime-Cam images. Of the 305 stars,
only 4 are misclassified as galaxies, which is a slightly lower
failure rate than the colour--colour selection of Caputi et
al.\ (2011) for this sample, while only 20 of 3116 galaxies are
wrongly classified as stars. Fig.~\ref{fig:bayes} shows histograms of
the probability ratio of Equation~\ref{eqn:bayes} for
spectroscopically-confirmed stars and galaxies. We further test our
classification method using the sample of objects that Hartley et
al.\ (2012) believe to be stars based on a variety of colour criteria
and, if bright enough, spatial profiles. Of the 2466 objects in common
with our catalogue, only 27 are (mis)classified as galaxies. We
therefore consider our star--galaxy classification to be $\ga99$\,per
cent reliable and as accurate as other methods. This results in a
sample of 96\,551 objects that are either spectroscopically classified
as galaxies or have no spectroscopic information but are considered
more likely to be galaxies from Equation~\ref{eqn:bayes}. The median
value of $\chi^2_{\rm g}$ for these objects is 7.8, and only 1\,350
(1.4\,per cent) have $\chi^2_{\rm g}>100$.

\begin{figure}
\resizebox{\hsize}{!}{\includegraphics{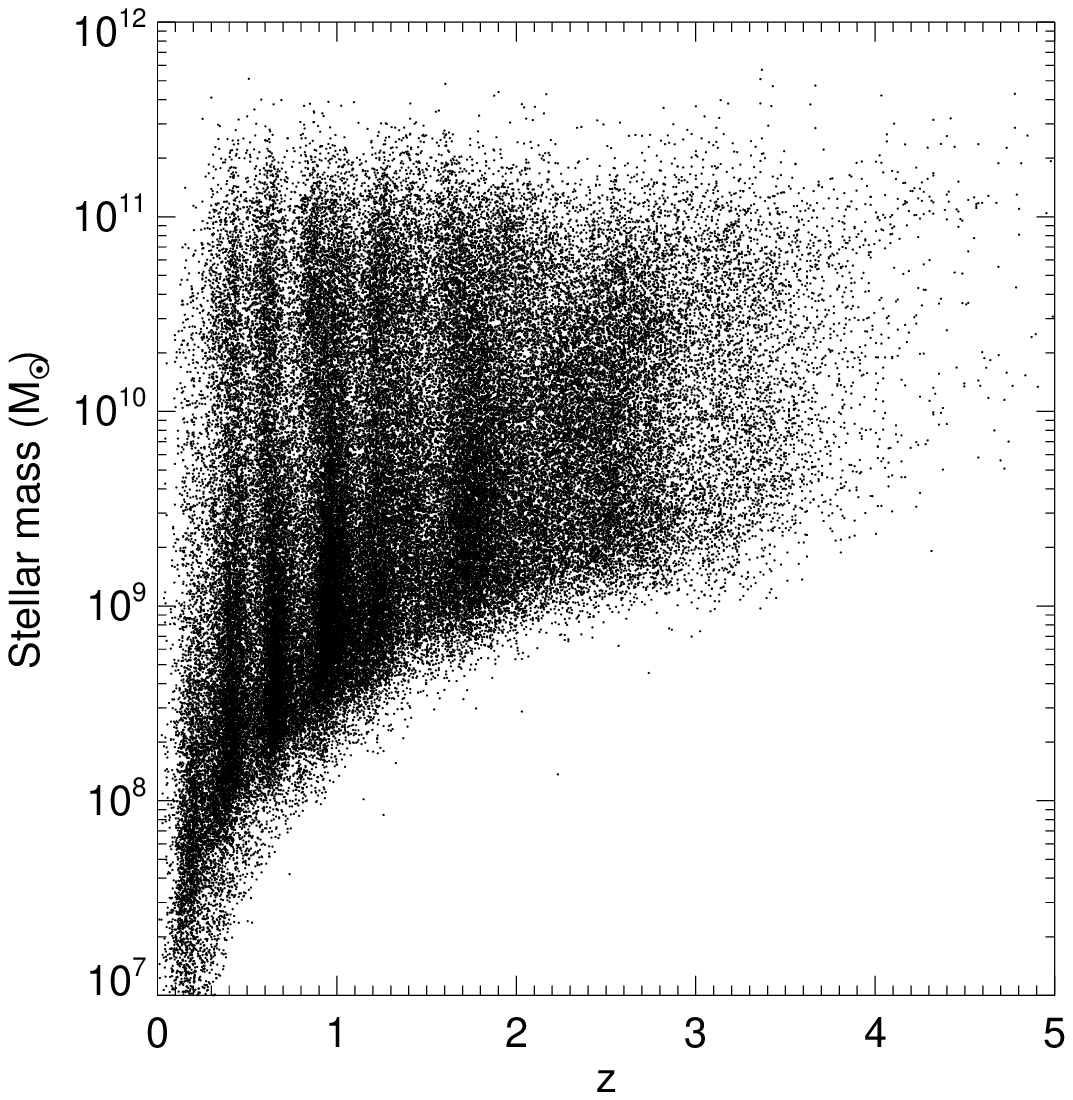}}
\caption[]{Plot of stellar mass against redshift for all sources
  classified as galaxies.}
\label{fig:masses}
\end{figure}

\begin{figure*}
\resizebox{\colwidth}{!}{\includegraphics{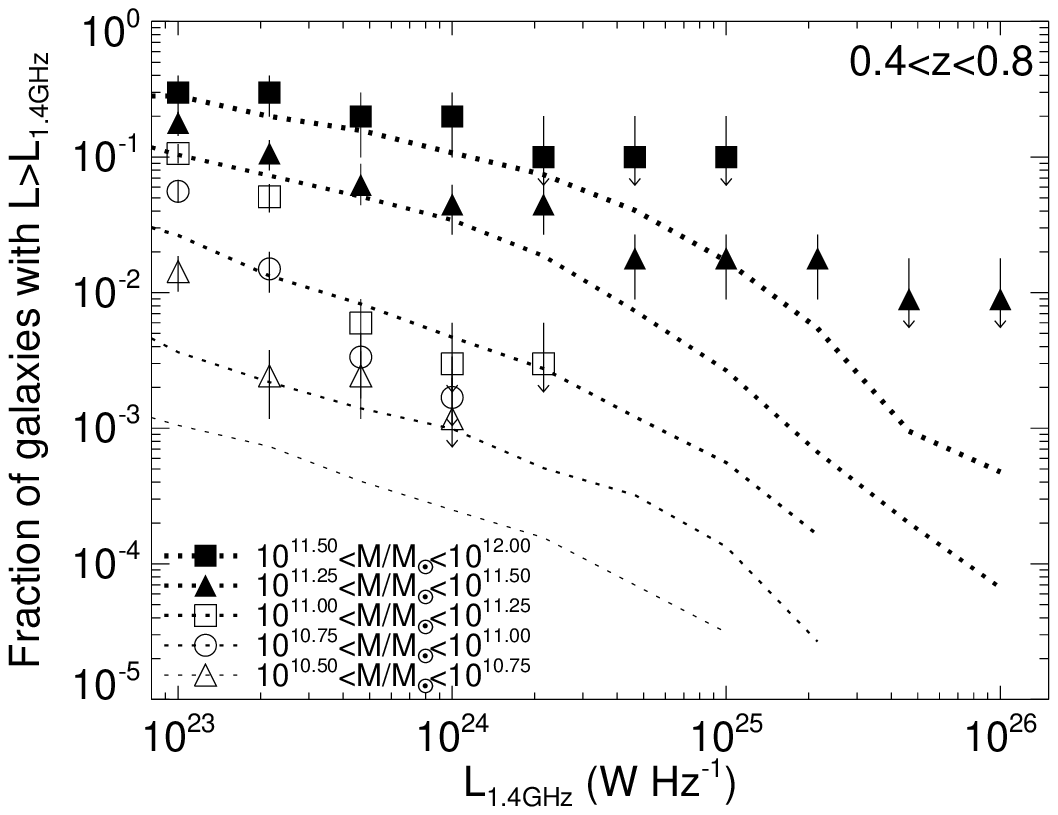}}
\resizebox{\colwidth}{!}{\includegraphics{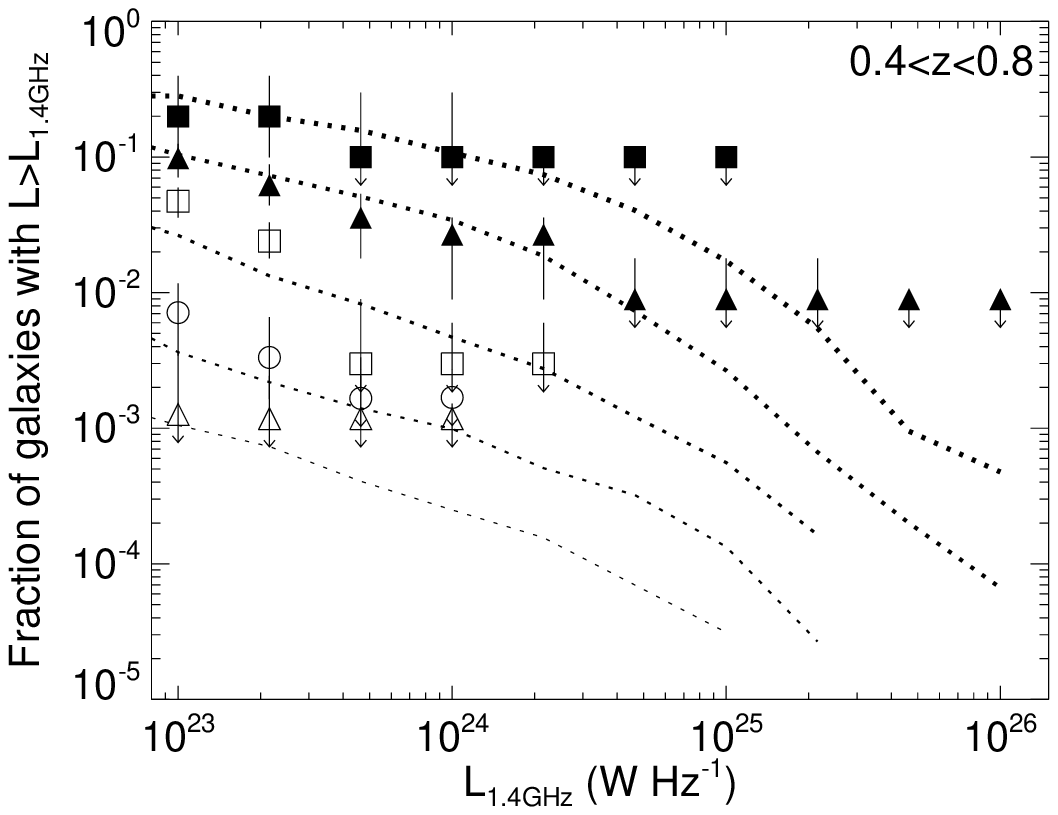}}
\resizebox{\colwidth}{!}{\includegraphics{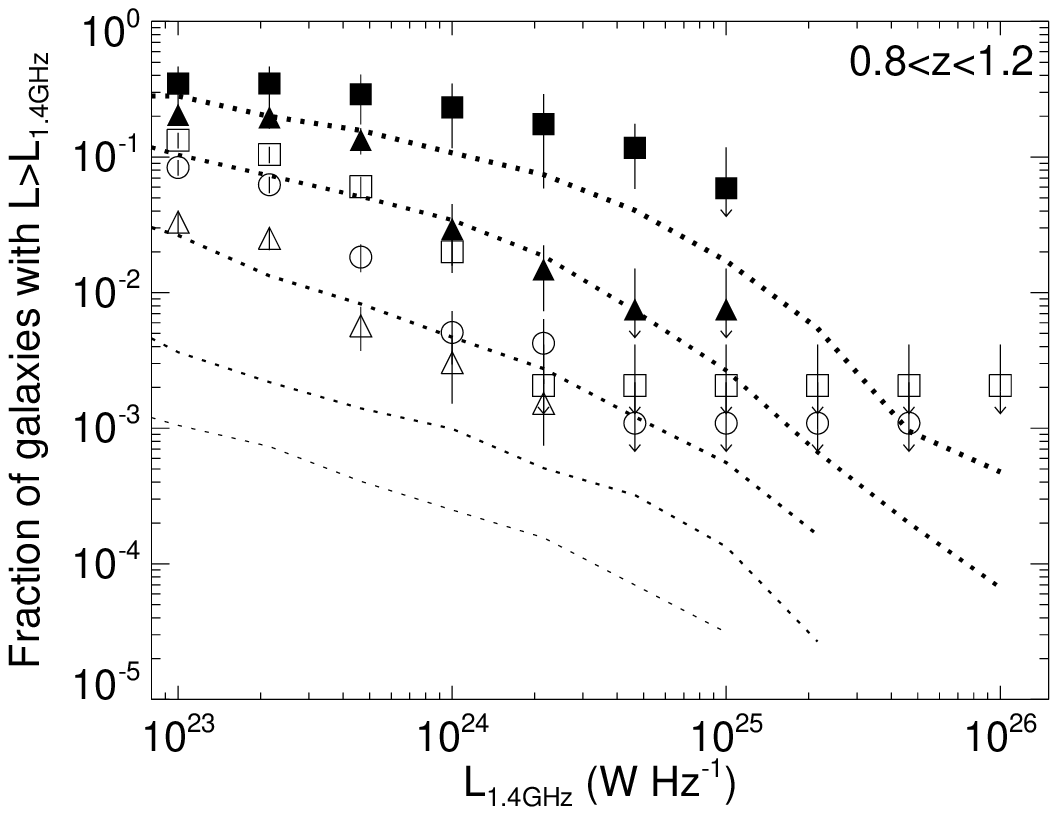}}
\resizebox{\colwidth}{!}{\includegraphics{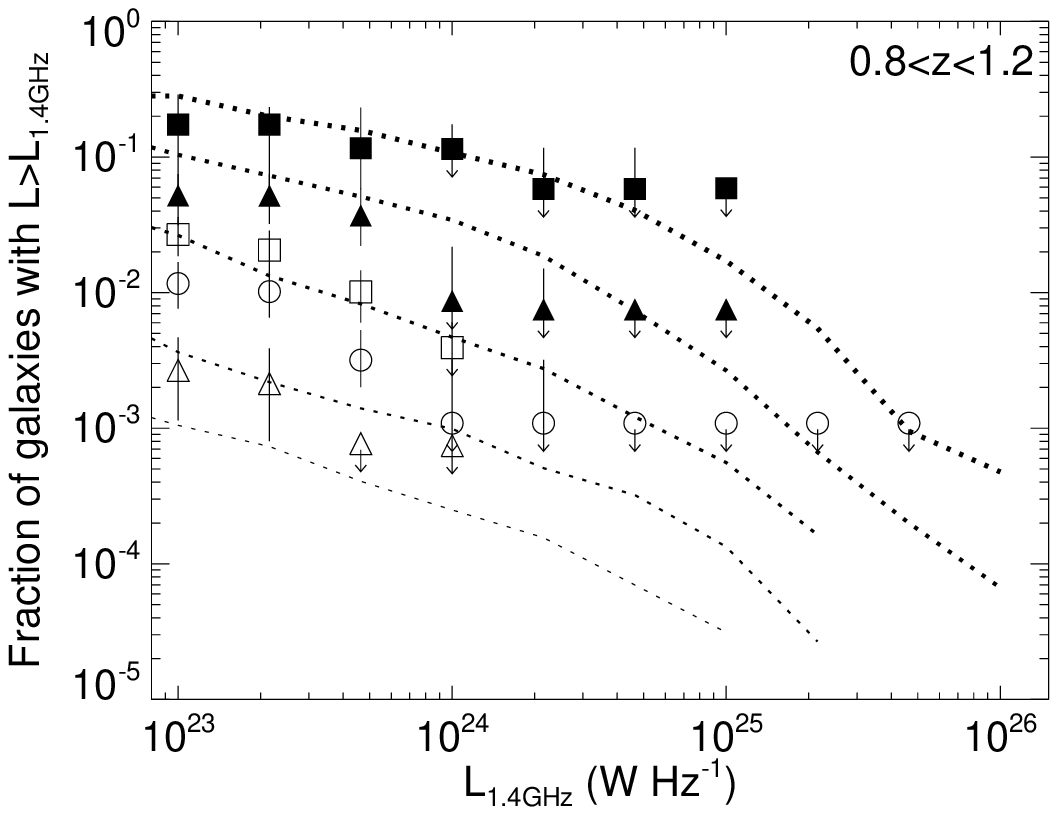}}
\caption[]{The fraction of galaxies with radio luminosities brighter
  than a given radio luminosity, in two redshift ranges, $0.4<z<0.8$
  (top) and $0.8<z<1.2$ (bottom). The left panels show all galaxies,
  while the right panels show the estimated fraction of hot-mode AGN,
  as described in the text. The data and uncertainties have been
  calculated by jackknifing, with the data points representing the
  median fraction and the error bars the central 68\,per cent
  confidence interval (with a downward-pointing arrow where this
  interval includes samples where the fraction is zero). Where the
  median fraction is zero but the confidence interval includes
  non-zero fractions, the data point is plotted at the upper
  confidence limit, with a downward-pointing arrow. No point is
  plotted when all fractions within the confidence interval are
  zero. The dotted lines show the data from Best et al.\ (2005b), and
  the key in the top-left figure applies to all plots. Our stellar
  masses have been increased by 13\,per cent to account for the
  different IMFs used.\label{fig:agnfracs}}
\end{figure*}

\section{Analysis and Discussion}

Fig.~\ref{fig:masses} shows a plot of stellar mass against redshift
for the galaxies in our sample. Some structure is seen at high stellar
masses as the result of peaks in the photometric redshift histogram,
including one at $z\approx0.65$ that coincides with a large-scale
structure possibly seen by Geach et al.\ (2007). We use these data to
construct samples of galaxies within redshift and stellar-mass limits
and investigate the incidence of radio emission within these samples.

We first investigate the issue of thermal balance in galaxies by
estimating the fraction of galaxies with radio luminosities above
various thresholds, in bins of stellar mass, following Best et
al.\ (2005b). Their stellar masses were taken from Kauffmann et
al.\ (2003), who used the same stellar library as BC03, but a Kroupa
(2001) IMF.  This is not available in the standard BC03 distribution,
with the Chabrier (2003) IMF being the closest option. Longhetti \&
Saracco (2009) calculate that the Kroupa IMF has a 13\,per cent higher
mass-to-light ratio than the Chabrier IMF and we apply this offset to
our stellar masses in order to allow a direct comparison with the
results of Best et al.\ (2005b, 2006).

To derive the number of galaxies within a given stellar mass and
redshift range whose radio luminosities exceed a given value, $L$, we
calculate the following, which incorporates our uncertainty in the
$K$-band counterparts and the incompleteness at faint radio flux
densities:
\begin{equation}
N(>L) = \sum_{k \in \{z,M\}} \sum_{r} P_{kr} / C(S_r) \, .
\end{equation}
Here $C(S_r)$ is the completeness at a flux density $S_r$ and the
second summation is over all radio sources whose flux densities make
them sufficiently luminous if their counterpart is the $K$-band source
being considered, i.e., if $4\pi d_{\rm
  l}(z_k)^2(1+z_k)^{\alpha-1}S_r>L$ where $d_{\rm l}(z_k)$ is the
luminosity distance for the source's redshift, $z_k$, and $\alpha=0.7$
is the assumed radio spectral index (in the sense $S_\nu \propto
\nu^{-\alpha}$). Since the number of $K$-selected galaxies is much
larger than the number of radio sources, most galaxies are not
associated with any radio source and no galaxy has a significant
probability of being associated with more than one radio
source. Therefore, in practice, the above equation reduces to a single
summation over all galaxies whose possible radio counterparts are
sufficiently bright.

We follow Best et al.\ (2005b) by calculating the fraction of galaxies
within a particular redshift and stellar mass range that host an AGN
above a given radio luminosity. We take 10\,000 jackknife samples of
the galaxies in each redshift--mass bin and sort them in order of AGN
fraction, using the 1587th and 8413rd fractions as estimates of the
1-$\sigma$ uncertainty. The results of this analysis in two redshift
intervals ($0.4<z\le0.8$ and $0.8<z\le1.2$) are shown in the left-hand
panels of Fig.~\ref{fig:agnfracs}, and compared to the fractions found
by Best et al.\ in their $z\le0.1$ sample. Our fractions are
systematically higher than those seen locally, and this difference is
especially pronounced at lower stellar masses
($M\la10^{11}$\,M$_\odot$) and/or in the higher redshift
interval. This is readily understood as being caused by the strong
evolution of both star-forming galaxies and radio-quiet AGN which,
while only important locally at radio luminosities
$L\la10^{23}$\,W\,Hz$^{-1}$, become the dominant population at
luminosities an order of magnitude higher by $z\sim1$ (Padovani et
al.\ 2011; S12).

Our interest lies with the hot-mode AGN and it is therefore necessary
to isolate these objects from the other classes of radio-emitting
extragalactic sources. Such objects can be identified by their weak or
absent emission lines but only a few per cent of our galaxy sample
have a sufficiently high-quality spectrum, and this subset is quite
strongly biased towards brighter, and hence lower-redshift, sources.
Mid-infrared imaging can be used to distinguish radio-loud AGN from
other sources via the $q_{24} = \log (S_{24\mu\rm m}/S_{\rm1.4GHz})$
statistic (Appleton et al.\ 2004; Ibar et al.\ 2008), but the
\textit{Spitzer\/} imaging is not deep enough to reliably discriminate
for sources with $S_{\rm1.4GHz}<100\,\mu$Jy (see fig.~6 of S12).
Furthermore, this statistic only determines whether a source is
radio-loud, not whether it is a hot-mode or cold-mode AGN.

Instead we use the parameters of the best-fitting stellar population
to infer the nature of the radio emission. S12 used such a method,
based on the fraction of rest-frame $V$-band light from young stars in
the EAzY spectral energy distribution. Since we are now using an
improved set of template spectra, we cannot adopt the exact method of
S12. After extensive experimentation, we adopt a rather more complex
approach that incorporates our uncertainty as to the true nature of
individual sources. However, we stress that our final results and
conclusions are robust to the exact methodology, provided we properly
account for the classification uncertainty.

We take all 253 radio sources with reliable spectroscopic redshifts
and classify these as weak-lined or strong-lined depending on whether
the rest-frame equivalent width of the
[O{\sc~ii}]~$\lambda\lambda$3726,3729 doublet is less than, or greater
than 5\,\AA, respectively. All sources lacking the spectral coverage
to cover [O{\sc~ii}] are classified as strong-lined as their redshifts
have been determined based on the presence of broad and/or
high-ionization emission lines that would not be present in truly
weak-lined objects. This is a tighter criterion than that used by S12
but ensures that no objects classified as narrow-line AGN or
star-forming galaxies on the basis of line ratios would have been
classified as weak-lined if the spectral coverage had only included
[O{\sc~ii}].

\begin{figure}
\resizebox{\hsize}{!}{\includegraphics{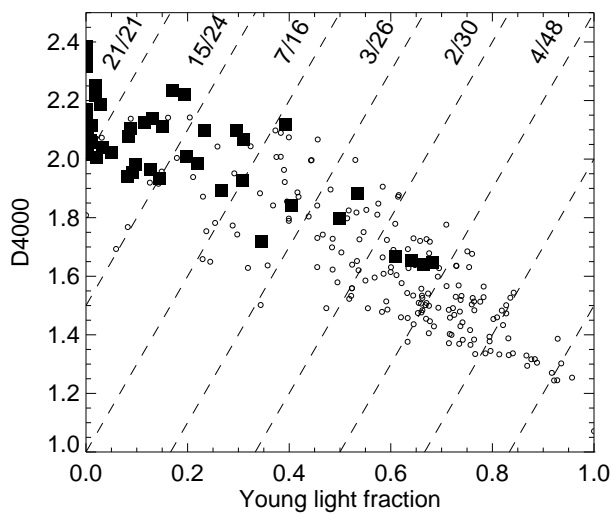}}
\caption[]{Plot of the 4000-\AA\ break strength, D4000, against the
  fraction of light from young stars at a rest-frame wavelength of
  3000\,\AA\ for all radio sources with spectra of sufficient quality
  to reliably identify them as weak-lined radio galaxies (filled
  squares) as opposed to star-forming galaxies or other AGN types
  (small open circles). This plane is split into regions as indicated
  by the dashed lines, with the number of such objects and the total
  number in each region indicated at the top of the figure.}
\label{fig:classify}
\end{figure}

For each galaxy, we use the composite stellar population from which
the stellar mass was derived to calculate the strength of the
4000-\AA\ break, D4000 (Bruzual 1983), and the fraction of light at a
rest-frame wavelength of 3000\,\AA\ that comes from stars with ages
younger than 250\,Myr. These values are displayed in
Fig.~\ref{fig:classify}, which shows the expected anti-correlation
between them, and the clear tendency for weak-lined sources to have
older stellar populations. The plane is divided into regions by
straight lines with a gradient of 3 (chosen to be perpendicular to the
locus of points) and the numbers of weak- and strong-lined sources
counted in each region. The fraction of weak-lined galaxies decreases
in a nearly monotonic fashion across these regions, and we use the
location of a galaxy in this figure to estimate the probability that
it is weak-lined.

\begin{figure*}
\resizebox{\colwidth}{!}{\includegraphics{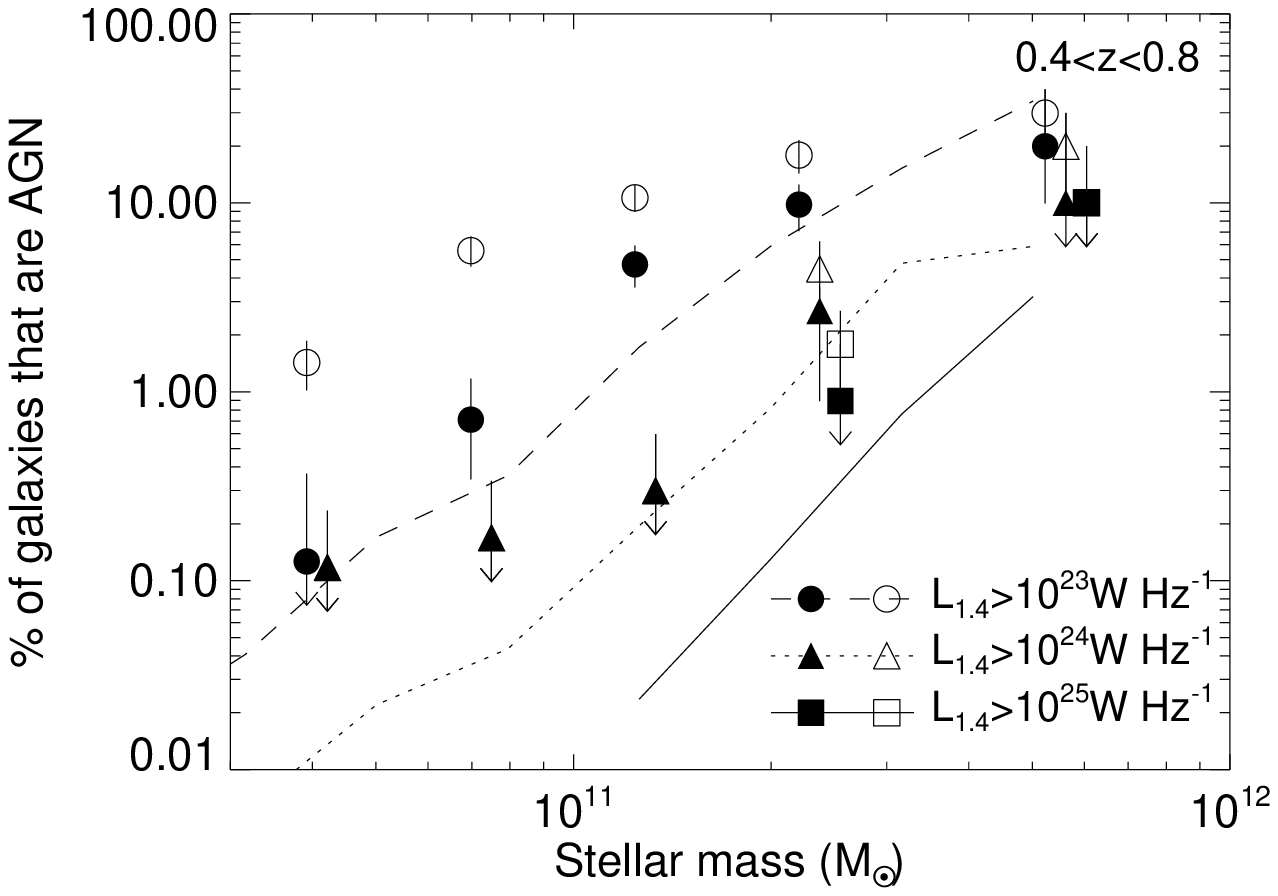}}
\resizebox{\colwidth}{!}{\includegraphics{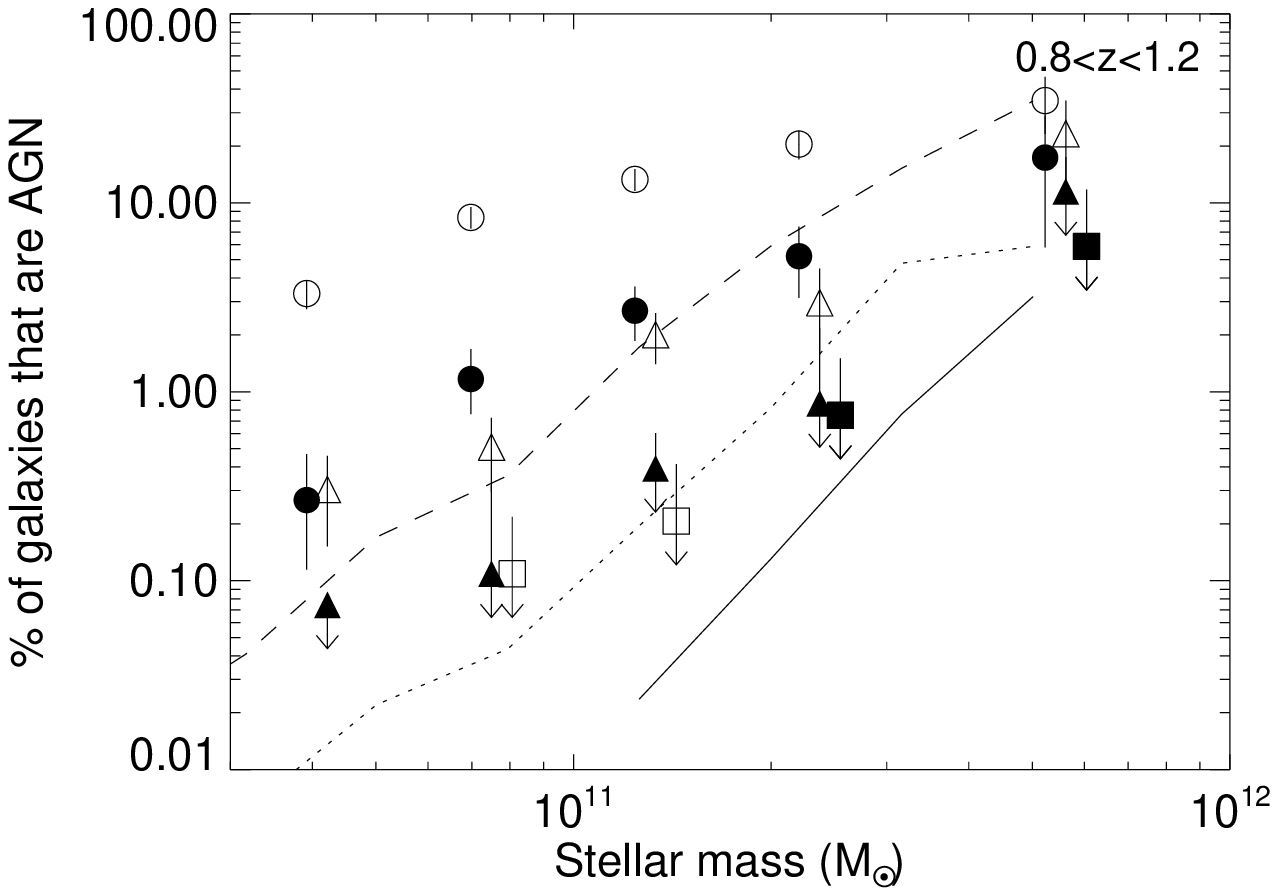}}
\caption[]{The fraction of galaxies with radio luminosities brighter
  than a limiting value, as a function of stellar mass. In each panel
  open symbols indicate the total fraction of galaxies, while filled
  symbols represent only the radio-loud fraction. These are the same
  data that are plotted in Fig.~\ref{fig:agnfracs}. The lines are
  taken from fig.~2 of Best et al.\ (2005b) and the horizontal
  offsets between points in the same mass bin are for clarity
  only. \label{fig:agnfracs2}}
\end{figure*}

We then repeat the analysis to estimate the fraction of galaxies
within each stellar mass--redshift bin that have a radio luminosity
greater than some limiting value, although this time it is modified to
estimate the fraction of weak-lined radio galaxies only. Since there
are few sources in each subdivision of Fig.~\ref{fig:classify}, the
true fraction of weak-lined radio sources in each region is not
strongly constrained.  We therefore make an estimate of the true
weak-lined galaxy fraction in each of the five regions where this is
neither 0 nor 1 before each jackknife run. If $m$ out of $n$ galaxies
are weak-lined, then the probability distribution function for the
true fraction of weak-lined galaxies, $f$, is given by
\begin{equation}
P(f | m,n) = \frac{n!}{m!(n-m)!} \, f^m (1-f)^{n-m}
\end{equation}
and this is used to estimate the weak-lined galaxy fractions. Then,
every time a galaxy with a sufficiently high radio luminosity is
selected during a jackknife run, a uniform random number in the
interval (0,1) is produced and only if this number is less than the
value of $f$ in that galaxy's region of Fig.~\ref{fig:classify} is the
galaxy then included.

The results of this analysis, which should provide the fraction of
hot-mode radio-loud AGN, are shown in the right-hand panels of
Fig.~\ref{fig:agnfracs}. Obviously the fraction is consistently lower
when just the weak-lined galaxies are considered, instead of the
entire AGN population. In the redshift interval $0.4<z\le0.8$ the
distribution of radio luminosities is entirely consistent with that
seen at $z<0.1$ by Best et al.\ (2005b), while there is tentative
evidence for a larger fraction of radio-loud AGN among the lower mass
galaxies in the $0.8<z\le1.2$ redshift bin, especially at radio
luminosities near our flux limit. Despite the small numbers of
galaxies in the highest mass bins, the consistency between the
estimated fractions in all three redshift intervals suggests that
galaxies of a given stellar mass have the same duty cycle of
radio-loud AGN activity across the entire redshift interval $0\leq z
\la 1$. This is also seen in Fig.~\ref{fig:agnfracs2}, where the AGN
fraction is displayed as a function of stellar mass for different
limiting radio luminosities. The strong evolution in the population of
luminous radio-quiet sources reported by Padovani et al.\ (2011) and
S12 is also apparent in this figure.

Our results do not necessarily imply that the balance between AGN
heating and radiative cooling inferred locally by Best et al.\ (2006)
also existed out to $z\sim1$, however. We note that the relationship
between radio luminosity and mechanical power for FR\,I radio sources
has been a subject of debate, with Cavagnolo et al.\ (2010) suggesting
a much steeper relationship than that of B\^{\i}rzan et al.\ (2004)
that is consistent with the model of Willott et al.\ (1999). This
actually has little effect on the mean heating rate since a steeper
relationship increases the importance of the rare, luminous sources
($L_{\rm1.4GHz}\ga10^{24}$\,W\,Hz$^{-1}$) towards the total energy
budget and the two relationships imply similar mechanical powers in
this range. The relationship between radiative cooling and stellar
mass used by Best et al.\ (2006) was derived from a sample of nearby
early-type galaxies but studies using deep X-ray data from
\textit{Chandra\/} have failed to find any significant evolution in
this relationship over the redshift range we are studying (Lehmer et
al.\ 2007; Danielson et al.\ 2012). It is therefore reasonable to
conclude that the balance between radiative cooling and AGN heating
had already been struck when the Universe was less than half its
current age. This balance maintains the temperature, and hence the
cooling rate, of the gaseous halo, and consequently it is to be
expected that the relationships between cooling rate and stellar mass,
and between radio luminosity and mechanical heating energy are the
same at $z\sim1$ as at the present epoch for galaxies that host
hot-mode AGN. 

\textbf{It is not a surprise that the relative locations of the points
  in different mass bins in Fig.~\ref{fig:agnfracs} agrees well with
  the local analysis, since both Best et al.\ (2005b) and Pope, Mendel
  \& Shabala (2012) show that equilibrium is maintained if the heating
  energy scales as $M_{\rm BH}^{1.5}$. However, we note that this does
  not require the shapes of the curves to be identical, since} the
same time-averaged heating rate could be achieved with a variety of
distributions of the frequency of activity as a function of radio
luminosity, e.g., with a high duty cycle of low-power activity or a
low duty cycle of higher luminosity activity. The agreement between
our data and the local relation of Best et al.\ (2005b) therefore
indicates not simply that equilibrium has been reached, but also that
the stochastic accretion events that maintain the equilibrium have the
same distribution of energies. The relative scalings between the

While the feedback mechanism is simple, it may not be initiated
immediately. Many scenarios for galaxy and black hole co-evolution
involve a final energetic episode of cold-mode accretion that sweeps
cold gas from the galaxy and returns it to the hot halo (e.g., Silk \&
Rees 1998; Fabian 1999). The cooling time for this newly-heated halo
depends on the amount of energy transferred to the gas and could
easily be of the order of 1\,Gyr or more, during which time the
accretion rate, and hence the heating rate, will be very low. There
will therefore be a delay between the final episode of cold-mode
activity and the instigation of a heating/cooling balance through
hot-mode feedback, and our result constrains any such episode to have
taken place well before $z\approx1$.

Our result differs from that of Danielson et al.\ (2012), who
estimated that the mechanical heating power from AGN exceeded the
radiative cooling losses by a factor of around 2 over the redshift
interval $0<z\la1.2$. This difference is readily explained by a
combination of two factors. First, Danielson et al.\ estimate the
mechanical heating power from \textit{all\/} AGN with radio
luminosities in the range $10^{22} < L_{\rm1.4GHz} / ({\rm
  W\,Hz}^{-1}) < 10^{24}$, irrespective of their nature. Their result
therefore includes the contribution from powerful radio-quiet sources
undergoing cold-mode accretion that should not be included when
balancing the radiative cooling in early-type galaxies. Second, these
objects are given significant weight in the calculation because the
authors adopt a much flatter relationship between mechanical power and
radio luminosity that boosts the relative contribution from
low-luminosity objects.

Since our data trace the curves of Best et al.\ (2005b) closely, and
these curves have not flattened off at our radio flux limit, there
must exist a significant population of hot-mode AGN even at radio flux
densities $S_{\rm1.4GHz} \ll 100\,\mu$Jy. Therefore, the common
practice of estimating the mean star-formation rate of an ensemble of
sources by stacking their radio images will overestimate this
quantity, even when individually-detected sources are removed (e.g.,
Simpson et al.\ 2006a; Dunne et al.\ 2009), although only by a modest
amount. For example, at $z\approx1$ only the very deepest radio
surveys have high completeness at
$L_{\rm1.4GHz}\approx10^{23}$\,W\,Hz$^{-1}$
($S_{\rm1.4GHz}\approx30\,\mu$Jy), which corresponds to a
star-formation rate of $\sim50$\,M$_\odot$\,yr$^{-1}$ in all stars
(Condon 1992; Kroupa 2001). Yet the curves of Best et al.\ (2005b)
suggest that $\sim10$\,per cent of massive galaxies host radio-loud
AGN with luminosities within a factor of 10 of this limit, which
equates to an overestimate of the mean star-formation rate by
$\sim2$\,M$_\odot$\,yr$^{-1}$. For shallower surveys where galaxies
with radio luminosities above this limit remain undetected, the effect
is greater.

Our results provide quantitative support for a strong causal link
between an absence of ongoing star formation in a galaxy and the
presence of a hot-mode AGN. Unfortunately, our study is unable to
determine whether such a link is still present at $z\ga1.5$, where the
colour bimodality in the galaxy population breaks down (Cirasuolo et
al.\ 2007), and where star formation in the most massive galaxies is
predicted to be curtailed by hot-mode feedback (Bower et al.\
2006). This is not due to the depth of the optical/infrared imaging
data, which Fig.~\ref{fig:masses} demonstrates reaches far lower
stellar masses than we are studying (although the high signal-to-noise
ratio photometry does result in more reliable stellar masses), but is
simply a result of the low number of extremely massive galaxies in the
UDS field. This problem can be overcome through the addition of other
fields that possess multi-band optical/infrared images and deep radio
mapping to $S_{\rm1.4GHz}<100\,\mu$Jy. We are currently undertaking a
study in the COSMOS/UltraVISTA field (Scoville et al.\ 2007;
Schinnerer et al.\ 2010; McCracken et al.\ 2012) and will obtain new
rest-frame optical spectroscopy to confirm and improve our selection
of hot-mode AGN, which remains a significant source of uncertainty.

\section{Summary}

We have described a new method for photometrically separating stars
and galaxies and deriving photometric redshifts and stellar masses,
and demonstrated its reliability and effectiveness using the extensive
spectroscopy in the UKIDSS Ultra Deep Survey field. Using this method,
we have derived redshifts and stellar masses for 96\,551 galaxies in
this field.

We have determined the distribution of radio luminosities among
galaxies in different redshift and stellar mass bins, and have shown
that, when only hot-mode AGN are considered, these distributions are
similar to those determined in the local Universe. This has been
interpreted as evidence for the energy balance between radiative
cooling and AGN heating in passive galaxies existing at least as least
as far back as 8\,Gyr. We suggest that existing data may cover a
sufficient volume to determine whether this balance existed a further
2\,Gyr earlier, before the colour bimodality in galaxies had existed.

\section*{Acknowledgments}

The authors thank the United Kingdom Science and Technology Facilities
Council for funding. We also wish to thank Ruth Gr\"{u}tzbauch for
providing a photometric redshift catalogue that was used in an early
version of this work, Maurizio Salaris for introducing us to France
Allard's library of stellar spectra, and Ian McCarthy for informative
discussions on the development of semi-analytic models and
simulations.

\label{lastpage}


\begin{thebibliography}{}

\bibitem{}Allard F., Homeier D., Freytag B., 2011, in Johns-Krull
  C. M., Browning M. K., West A. A., eds ASP Conf.\ Ser.\ Vol.~448,
  16th Cambridge Workshop on Cool Stars, Stellar Systems, and the
  Sun. Astron.\ Soc.\ Pac., San Francisco, p.~91

\bibitem{}Appleton P. N., et al., 2004, ApJS, 154, 147

\bibitem{}Becker R. H., White R. L., Helfand D. J., 1995, ApJ, 450, 559

\bibitem{}Bennett A. S., 1962, MemRAS, 68, 163

\bibitem{}Bertin E., Arnouts S., 1996, A\&AS, 117, 393

\bibitem{}Best P. N., Kaiser C. R., Heckman T. M., Kauffmann G., 2006,
  MNRAS, 368, L67

\bibitem{}Best P. N., Kauffmann G., Heckman T. M., Ivezi\'{c} \v{Z},
  2005a, MNRAS, 362, 9

\bibitem{}Best P. N., Kauffmann G., Heckman T. M., Brinchmann J.,
Charlot S., Ivezi\'{c} \v{Z}., White S. D. M., 2005b, MNRAS, 362, 25

\bibitem{}B\^{\i}rzan L., Rafferty D. A., McNamara B. R., Wise M. W.,
  Nulsen P. E. J., 2004, ApJ, 607, 800

\bibitem{}B\"{o}hringer H., Voges W., Fabian A. C., Edge A. C.,
  Neumann D. M., 1993, MNRAS, 264, L25

\bibitem{}Bondi M., Ciliegi P., Schinnerer E., Smol\v{c}i\'{c} V.,
  Jahnke K., Carilli C., Zamorani G., 2008, ApJ, 681, 1129

\bibitem{}Booth C. M., Schaye J., 2009, MNRAS, 398, 53

\bibitem{}Bower R. G., Benson A. J., Malbon R., Helly J. C., Frenk
  C. S., Baugh C. M., Cole S., Lacey C. G., 2006, MNRAS, 370, 645

\bibitem{}Bower R. G., McCarthy I. G., Benson A. J., 2008, MNRAS, 390,
  1399

\bibitem{}Brammer G. B., van Dokkum P. G., Coppi P., 2008, ApJ, 686,
  1503

\bibitem{}Bruzual G., 1983, ApJ, 273, 105

\bibitem{}Caputi K. I., Cirasuolo M., Dunlop J. S., McLure R. J.,
  Farrah D., Almaini O., 2001, MNRAS, 413, 162

\bibitem{}Cattaneo A., Dekel A., Devriendt J., Guiderdoni B., Blaizot
  J., 2006, MNRAS, 370, 1651 

\bibitem{}Cavagnolo K. W., McNamara B. R., Nulsen P. E. J., Carilli
  C. L., Jones C., B\^{\i}rzan L., 2010, ApJ, 720, 1066

\bibitem{}Chabrier G., 2003, PASP, 115, 763

\bibitem{}Chabrier G., Baraffe I., Leconte J., Gallardo J., Barman T.,
  2009, Stempels E., ed.\ AIP Conf.\ Proc.\ Vol.~1094, 15th Cambridge
  Workshop on Cool Stars, Stellar Systems and the Sun. AIP, New York,
  p.~102 (arXiv:0810.5085v1)

\bibitem{}Cirasuolo M., et al., 2007, MNRAS, 380, 585

\bibitem{}Clewley L., Jarvis M. J., 2004, MNRAS, 352, 909

\bibitem{}Condon J. J., 1984, ApJ, 284, 44

\bibitem{}Condon J. J., 1992, ARA\&A, 30, 575

\bibitem{}Condon J. J., 1997, PASP, 109, 166

\bibitem{}Condon J. J., Cotton W. D., Greisen E. W., Yin Q. F., Perley
  R. A., Taylor G. B., Broderick J. J., 1998, AJ, 115, 1693

\bibitem{}Croton D. J., et al., 2006, MNRAS, 365, 11

\bibitem{}Danielson A. L. R., Lehmer B. D., Alexander D. M., Brandt
  W. N., Luo B., Miller N., Xue Y. Q., Stott J. P., 2012, MNRAS, 422,
  494

\bibitem{}de Ruiter H. R., Arp H. C., Willis A. G., 1977, A\&AS, 28, 211

\bibitem{}Demircan O., Kahraman G., 1991, Ap\&SS, 181, 313

\bibitem{}Dunne L., et al., 2009, MNRAS, 394, 3

\bibitem{}Dunlop J. S., Peacock J. A., 1990, MNRAS, 247, 19

\bibitem{}Eddington A. S., 1913, MNRAS, 73, 359

\bibitem{}Fabian A. C., 1999, MNRAS, 308, L39

\bibitem{}Fabian A. C., Sanders J. S., Allen S. W., Crawford C. S.,
  Iwasawa K., Johnstone R. M., Schmidt R. W., Taylor G. B., 2003,
  MNRAS, 344, L43

\bibitem{}Fanidakis N., Baugh C. M., Benson A. J., Bower R. G., Cole
  S., Done C., Frenk C. S., 2011, MNRAS, 410, 53

\bibitem{}Furusawa H., et al., 2008, ApJS, 176, 1

\bibitem{}Gardner J. P., Cowie L. L., Wainscoat R. J., 1993, ApJ, 415,
  L9

\bibitem{}Geach J. E., Simpson C., Rawlings S., Read A. M., Watson M.,
  2007, MNRAS, 381, 1369

\bibitem{}Hartley W. G., et al., 2013, MNRAS, submitted

\bibitem{}Hopkins A. M., Afonso J., Chan B., Cram L. E., Georgakakis
  A., Mobasher B., 2003, AJ, 125, 465

\bibitem{}Ibar E., et al., 2008, MNRAS, 386, 953

\bibitem{}Ibar E., Ivison R. J., Biggs A. D., Lal D. V., Best P. N.,
  Green D. A., 2009, MNRAS, 397, 281

\bibitem{}Ivison R. J., et al., 2007, MNRAS, 380, 199

\bibitem{}Jarrett T. H., 1992, Ph.D.\ Thesis, University of
  Massachusetts

\bibitem{}Kauffmann G., et al., 2003, MNRAS, 341, 33

\bibitem{}Komatsu E., et al., 2011, ApJS, 192, 18

\bibitem{}Kroupa P., 2001, MNRAS, 322, 231

\bibitem{}Laing R. A., Riley J. M., Longair M. S., 1983, MNRAS, 204,
  151

\bibitem{}Lawrence A., et al., 2007, MNRAS, 379, 1599

\bibitem{}Longhetti M., Saracco P., 2009, MNRAS, 394, 774

\bibitem{}Leggett S. K., et al., 2010, ApJ, 710, 1627

\bibitem{}Lehmer B. D., et al., 2007, ApJ, 657, 681

\bibitem{}McAlpine K., Jarvis M. J., 2011, MNRAS, 413, 1054

\bibitem{}McCarthy I. G., et al., 2010, MNRAS, 406, 822

\bibitem{}McCracken H. J., et al., 2012, A\&A, 544, 156

\bibitem{}Padovani P., Miller N., Kellermann K. I., Mainieri V.,
  Rosati P., Tozzi P., 2011, ApJ, 740, 20

\bibitem{}Pei Y. C., 1992, ApJ, 395, 130

\bibitem{}Pope E. C. D., 2011, MNRAS, 414, 3344

\bibitem{}Popesso P., et al., 2009, A\&A, 494, 443

\bibitem{}Puchwein E., Sijacki D., Springel V., 2008, ApJ, 687, L53

\bibitem{}Reichardt C., Jiminez R., Heavens A. F., 2001, MNRAS, 327,
  849

\bibitem{}Reid M. J., Schneps M. H., Moran J. M., Gwinn C. R., Genzel
  R., Downes D., R\"{o}nn\"{a}ng B., 1988, AJ, 116, 1039

\bibitem{}Rigby E. E., Best P. N. Brookes M. H., Peacock J. A., Dunlop
  J. S., R\"{o}ttgering H. J. A., Wall J. V., Ker L., 2011, MNRAS,
  416, 1900

\bibitem{}Santini P., et al., 2012, A\&A, 538, A33

\bibitem{}Schinnerer E., et al., 2010, ApJS, 188, 384

\bibitem{}Scoville N., et al., 2007, ApJS, 172, 1

\bibitem{}Shabala S. S., Ash S., Alexander P., Riley J. M., 2008,
  MNRAS, 388, 625

\bibitem{}Sijacki D., Springel V., Di Matteo T., Hernquist L., 2007,
  MNRAS, 380, 877

\bibitem{}Silk J., Rees M. J., 1998, A\&A, 331, L1

\bibitem{}Simpson C., et al., 2006a, MNRAS, 372, 741 (S06)

\bibitem{}Simpson C., et al., 2006b, MNRAS, 373, L21

\bibitem{}Simpson C., et al., 2012, MNRAS, 421, 3060 (S12)

\bibitem{}Sivia D. S., 1996, Data Analysis: A Bayesian
  Tutorial. Oxford Univ.\ Press, Oxford

\bibitem{}Smol\v{c}i\'{c} V., et al., 2008, ApJS, 177, 14

\bibitem{}Sutherland W., Saunders W., 1992, MNRAS, 259, 413

\bibitem{}Vermeulen R. C., 2012, Stepp L. M., Gilmozzi R., Hall H. J.,
  eds SPIE~8444, Ground Based and Optical Telescopes IV, 2B

\bibitem{}Waddington I., Dunlop J. S., Peacock J. A., Windhorst R. A.,
  2001, MNRAS, 328, 896

\bibitem{}Willott C. J., Rawlings S., Blundell K. M., Lacy M., 1999,
  MNRAS, 309, 1017

\bibitem{}Wilman R. J., et al., 2008, MNRAS, 388, 1335

\bibitem{}Windhorst R. A., Miley G. K., Owen F. N., Kron R. G., Koo
  D. C., 1985, ApJ, 289, 494

\bibitem{}Windhorst R. A., Dressler A., Koo D. C., 1987, Hewitt A., et
  al., eds Observational Cosmology, IAU, p.~573

\bibitem{}Wolstencroft R. D., Savage A., Clowes R. G., MacGillivray
  H. T., Leggett S. K., Kalafi M., 1986, MNRAS, 223, 279

\bibitem{}York D. G., et al., 2000, AJ, 120, 1579

\end{thebibliography}
\end{document}